\documentclass[12pt]{article}

\ifx\pdfoutput\undefined
\usepackage[dvips,bookmarks]{hyperref}  
\else
\usepackage{hyperref}   
\fi

\usepackage{graphicx}

\hypersetup{colorlinks,bookmarksopen,bookmarksnumbered,citecolor=blue,pdfstartview=FitH}
\def\hhref#1{\href{http://arxiv.org/abs/hep-th/#1}{hep-th/#1}} 
\def\mhref#1{\href{Milton:#1}{#1}}              

\def\bop#1{\setbox0=\hbox{$#1M$}\mkern1.5mu
\vbox{\hrule height0pt depth.04\ht0
\hbox{\vrule width.04\ht0 height.9\ht0 \kern.9\ht0
\vrule width.04\ht0}\hrule height.04\ht0}\mkern1.5mu}
\def\bo{{\mathpalette\bop{}}}                        
\def\frac#1#2{{#1\over#2}}     
\def\f#1#2{{\textstyle{#1\over#2}}}     
\def\ha{{\textstyle{1\over 2}}}     

\parskip=\medskipamount         
\thicklines             
\begin{document}

\newcommand{\be}{\begin{equation}}
\newcommand{\ee}{\end{equation}}
\newcommand{\mx}{\mbox}
\newcommand{\mt}{\mathtt}
\newcommand{\p}{\partial}
\newcommand{\st}{\stackrel}
\newcommand{\al}{\alpha}
\newcommand{\bb}{\beta}
\newcommand{\ga}{\gamma}
\newcommand{\te}{\theta}
\newcommand{\de}{\delta}
\newcommand{\et}{\gamma}
\newcommand{\ze}{\zeta}
\newcommand{\s}{\sigma}
\newcommand{\e}{\epsilon}
\newcommand{\om}{\omega}
\newcommand{\Om}{\Omega}
\newcommand{\la}{\lambda}
\newcommand{\La}{\Lambda}
\newcommand{\ti}{\widetilde}
\newcommand{\ih}{\hat{i}}
\newcommand{\jh}{\hat{j}}
\newcommand{\kh}{\widehat{k}}
\newcommand{\lh}{\widehat{l}}
\newcommand{\eh}{\widehat{e}}
\newcommand{\ph}{\widehat{p}}
\newcommand{\qh}{\widehat{q}}
\newcommand{\mh}{\widehat{m}}
\newcommand{\nh}{\widehat{n}}
\newcommand{\Dh}{\widehat{D}}
\newcommand{\2}{{\textstyle{1\over 2}}}
\newcommand{\3}{{\textstyle{1\over 3}}}
\newcommand{\4}{{\textstyle{1\over 4}}}
\newcommand{\8}{{\textstyle{1\over 8}}}
\newcommand{\6}{{\textstyle{1\over 16}}}
\newcommand{\ra}{\rightarrow}
\newcommand{\lra}{\longrightarrow}
\newcommand{\Ra}{\Rightarrow}
\newcommand{\im}{\Longleftrightarrow}
\newcommand{\hs}{\hspace{5mm}}
\newcommand{\bea}{\begin{eqnarray}}
\newcommand{\eea}{\end{eqnarray}}
\newcommand{\NP}{{\em Nucl.\ Phys.\ }}
\newcommand{\AP}{{\em Ann.\ Phys.\ }}
\newcommand{\PL}{{\em Phys.\ Lett.\ }}
\newcommand{\PR}{{\em Phys.\ Rev.\ }}
\newcommand{\PRL}{{\em Phys.\ Rev.\ Lett.\ }}
\newcommand{\PRP}{{\em Phys.\ Rep.\ }}
\newcommand{\CMP}{{\em Comm.\ Math.\ Phys.\ }}
\newcommand{\MPL}{{\em Mod.\ Phys.\ Lett.\ }}
\newcommand{\IJMP}{{\em Int.\ J.\ Mod.\ Phys.\ }}
\begin{titlepage}
\setcounter{page}{0} 
\begin{flushright}
YITP-SB-07-13\\July 10, 2007\\
\end{flushright}

\begin{center}
{\centering {\LARGE\bf
Running anti-de Sitter radius\\
from QCD-like strings

\par} }

\vskip 1cm {\bf Yu-tin Huang\footnote{E-mail address:
\mhref{yhuang@grad.physics.sunysb.edu}}, Warren Siegel\footnote{E-mail
address: \mhref{siegel@insti.physics.sunysb.edu}}} \\ \vskip 0.5cm

{\it C.N. Yang Institute for Theoretical Physics,\\
State University of New York, Stony Brook, 11790-3840 \\}

\end{center}

\begin{abstract}
We consider renormalization effects for a bosonic QCD-like string,
whose partons have $1/p^{2}$ propagators instead of Gaussian. Classically this model resembles (the bosonic part of) the projective light-cone (zero-radius) limit  of a string on an AdS${}_5$ background,
where Schwinger parameters give rise to the fifth dimension.
Quantum effects generate dynamics for this dimension,
producing an AdS${}_5$ background with a running radius.
The projective light-cone is the high-energy limit:
Holography is enforced dynamically.
\end{abstract}

\end{titlepage}

\section{Introduction}\label{1}

It has been argued that for a string theory to describe 4-dimensional gauge
theories (QCD strings) it must live in five dimensions \cite{polyakov}. For
the usual bosonic string theory outside of the critical dimension, the fifth
dimension arises from the conformal anomaly, the Liouville field. Consistent
quantization of Liouville theory that preserves the conformal symmetry is
essential for the understanding of non-critical strings and is still an
active research area.

Another way to understand the quantum string is to replace the
world-sheet by a random lattice:
The lattice is the Feynman diagram of ``partons" that compose
the string:  Each link is identified with a propagator, and the vertices are
the interaction vertices \cite{fishnet}.
The randomness of the lattice, corresponding to different geometries, is associated with the summation over different Feynman diagrams \cite{David}.
This approach was first applied to
understand pure 2D quantum gravity, and
in conjunction with the 1/N expansion (which defines ``planarity" for diagrams) \cite{Hooft},
the connection was made with the
Liouville approach in the continuum limit \cite{random}.

However, the bosonic (or super \cite{fs}) lattice
string has several unsatisfactory properties at large transverse momentum
for the underlying parton theory, such as Gaussian propagators and no
particle degrees of freedom in the deconfinement phase. In \cite{W. Siegel1} one introduces a Schwinger
parameter to give the usual $1/p^2$ propagators, which gives rise
to a QCD-like string that predicts the correct dimension 4 for preserving
T-duality. For such a theory the open string is identified as ``mesons" while
the closed string is ``pomerons". However, little success has been obtained in this
approach except for scalar partons.

The AdS/CFT correspondence also gives a correspondence between gauge theory
and string theory \cite{adscft}. The IIB string states correspond to
color-singlet bound states of N=4 super Yang-Mills. An important ingredient
is ``holography", which conjectures that the dynamical
properties are uniquely determined by the four-dimensional boundary theory.
The background AdS${}_5\otimes$S${}^5$ has isometries
SO(4,2) for AdS${}_5$ and SO(6) for S${}^5$, which are the same as the 4-dimensional
conformal group and the SU(4)${}_R$ of N=4 SYM.  It was shown in
\cite{Nastase:2000za} that by taking another limit, the projective
light-cone limit, one obtains a different holography where the fifth
dimension is still present, albeit non-dynamical to leading order. Random
latticizing this superstring it was shown that the bosonic part corresponds
to a wrong-sign $\phi^{4}$ theory similar to that used in \cite{W. Siegel1},
while the entire superstring gives a manifestly N=4 supersymmetric action
for a matrix field identified with N=4 SYM.

In this paper we take the previous QCD-like string \cite{W. Siegel1} and
perform a one-loop calculation.
A dynamical AdS radius is generated for a fifth dimension arising from the Schwinger parameter.
(This field already appears classically, and so is not the Liouville mode.)
This radius runs:  At high energy the theory is asymptotically free in this dynamically produced coupling, producing four-dimensional space as the projective light-cone limit \cite{Nastase:2000za}.

\section{QCD-like strings}\label{2}

In the usual random lattice quantization approach one expresses the string
world-sheet as a random lattice, using the irregularity of the lattice to represent world-sheet curvature:
\begin{equation}
\int DX\ e^{-S}\sim \int \prod_{i}dx_{i}\ e^{-\ha\sum_{\langle ij
\rangle}{(x_{i}-x_{j})}^{2}+\mu\sum_{i}1-\log N (\sum_{j}1-\sum_{\langle ij
\rangle}1+\sum_{I}1)}
\end{equation}
where $x_{i}$ are the vertices, $\langle ij \rangle$ label the links or
propagators and I are the faces. Thus summing over different lattices corresponds to
integrating over different geometries of the world-sheet. One can then
identify the lattice with the Feynman diagrams of some underlying (parton) field
theory. One can obtain the Feynman rules from the lattice string
action: (1) The usual $\ha(\partial X)^{2}$ term becomes on the
lattice $\ha(x_{i}-x_{j})^{2}$, giving a Gaussian propagator for the
parton theory. (2) The 2D cosmological term gives the world-sheet area and corresponds to the
number of vertices, and is thus related to the coupling constant. (3) The curvature
term has the usual interpretation of the $1/N$ expansion in the
parton theory.

The Gaussian propagators produce non-parton like behavior at large
transverse momenta \cite{Gaussian} and produce no degrees of freedom beyond the
Hagedorn temperature \cite{poles} (there are no poles in the propagator),
where there should be parton degrees of freedom in the deconfinement
phase. One can incorporate the usual $1/p^2$ in the random lattice
approach by using Schwinger parameters \cite{W. Siegel1}. That is, we can
write:
\begin{equation}
{1\over p^2}=\int_{0}^{\infty}d\tau\ e^{-\tau p^{2}}
\end{equation}
A Feynman diagram with non-derivative interactions can now be written in a
first-quantized form:
\begin{equation}
\int dp_{ij}dx_{i}d\tau_{ij}\ e^{-\ha\sum_{\langle ij
\rangle}[\tau_{ij}p_{ij}^{2}+i(x_{i}-x_{j})\cdot p_{ij}]}
\end{equation}
Integration over the vertices $x_{i}$ gives momentum conservation at each
vertex, while the $\tau$ integration gives the propagators. Taking the
underlying parton theory as wrong-sign $\phi^{4}$ theory, wrong sign meaning a negative coupling constant since the string amplitudes are always positive, then each vertex has two independent
propagators. Therefore in the continuum limit $\tau^{mn}$ has two components at
each point on the world-sheet which is a symmetric traceless tensor. This
suggests the following continuum action:
\begin{eqnarray}
L&=&\ha\tau_{mn}p^{m}\cdot
p^{n}+\lambda\tau^{mn}g_{mn}+ip^{m}\cdot\partial_{m}x+L_{g} \nonumber\\
L_{g}&=&\sqrt{-g}\left(\Lambda -R\ln N +\frac{c}{24}R{1\over\bo}R\right)
\end{eqnarray}
(where $\tau^{mn}$ is the inverse of $\tau_{mn}$).  $L_{g}$, which
depends only on the world-sheet metric, includes the cosmological constant and curvature terms, while $\lambda$ is the Lagrange
multiplier enforcing the traceless condition. Integrating out $p$,
\begin{equation}
L=\ha\tau^{mn}(\partial_{m}x)\cdot(\partial_{n}x)+\lambda\tau^{mn}g_{mn}+L_{g}
\end{equation}

The $R(1/\bo)R$ term was expected from quantum effects.  (It really belongs in the effective action; in the continuum case it comes from ghosts, but on the lattice the analog of ghosts is obscure.)  We can determine its coefficient by comparison with ordinary string theory:  In $D=0$ there is no $x$, and $\tau$ becomes irrelevant, so there the QCD-like string is identical to the usual string.  The metric $g_{mn}$ then describes simply the counting of Feynman diagrams, with respect to the $1/N$ expansion, with no dynamics.  But we know the continuum limit there:  It's the usual action for the $D=0$ subcritical string.  Thus, quantization of the metric will produce the usual $c=-26$ from the ghosts, which is now not canceled by $x$, which does not couple directly to the metric.
(We could also choose a gauge in terms of $\tau$ rather than $g$, which is more practical for the rest of the analysis, but then $g$ would be propagating and its one-loop evaluation more complicated.)
Such a term is necessary also because in its absence the equations of motion for $g$ and the constraint induced by
the Lagrange multiplier
\begin{equation}
\lambda\tau^{mn}=-\ha\Lambda\sqrt{-g}g^{mn},\quad\tau^{mn}g_{mn}=0
\end{equation}
are incompatible.  After choosing a gauge (in terms of either $g$ or $\tau$), this part of the theory totally decouples from the $x$
fields classically but comes in through a Liouville mode in the effective action to maintain local scale invariance as we will show.

Since there are no self-interactions in $x$, one-loop calculations give the
complete contribution of $x$ to the effective action in $\tau$.  In practice one first introduces vertex operators that depend only on $x$; integrating out $x$ then gives this $\tau$ action, as well as the usual factors of the $x$ Green function (now $\tau$-dependent) multiplying external-line momenta and polarizations.

\section{One loop integral}

We now compute the one-loop two-point integral for the $\tau$ field.
This will be sufficient to determine the contribution of $x$ to the renormalization and renormalization group behavior of the theory.
We assign the
vacuum expectation value $\langle\tau\rangle^{ab}$ for the tree-level $x$ propagator, and then restore an arbitrary $\tau$ background using 2D coordinate invariance.

In arbitrary world-volume dimension D (where on the world-sheet D=2) we calculate the 2-point effective action
$$
\Gamma [\tau] = \int \frac{d^{D}p}{(2\pi)^{D/2}} {\cal A} (\tau, p)
$$
\begin{equation}
{\cal A} = - \frac{d}{2}\int \frac{d^{D}k}{(2\pi)^{D/2}}
\tau^{ab}(p){(k+\ha p)_a (k-\ha p)_b (k+\ha p)_c (k-\ha p)_d
\over\langle\tau\rangle^{kl}(k+\ha p)_{k} (k+\ha p)_{l}\langle\tau\rangle^{mn}(k-\ha p)_{m}(k-\ha p)_{n}}\tau^{cd}(-p)
\end{equation}
With space-time dimension d=4 the integral gives:
\begin{eqnarray}
&&\frac{1}{4\sqrt{\langle\tau\rangle}}(\f18\langle\tau\rangle^{kl}p_{k}p_{l})^{D/2}{\Gamma(\ha)\over \Gamma(\frac{D+3}{2})}\Gamma(1-\f{D}{2})\Gamma(\f{D}{2}+1) \times
\nonumber\\
&&\Bigg\{ \f2D[(\tau^{ab}\langle\tau^{-1}_{ab}\rangle)^{2}+2\tau^{ac}\langle\tau^{-1}_{cd}\rangle\tau^{db}\langle\tau^{-1}_{ba}\rangle]
+\ha\tau^{ab}\langle\tau^{-1}_{ab}\rangle{p_c p_d\over \langle\tau\rangle^{kl}p_{k}p_{l}}\tau^{cd}
\nonumber\\
&&-\f1D\tau^{ab}\langle\tau^{-1}_{bc}\rangle {p_a p_d\over \langle\tau\rangle^{kl}p_{k}p_{l}}\tau^{cd}
-(1-\f{D}{2})\tau^{ab}\frac{4p_{a}p_{b}p_{c}p_{d}}{(\langle\tau\rangle^{kl}p_{k}p_{l})^2}\tau^{cd} \Bigg\}
\end{eqnarray}
with $\langle\tau\rangle=det\langle\tau^{ab}\rangle$.
Using $D=2+2\epsilon$ we arrive at
$$
{\cal A} (\tau, p) = \frac{1}{\sqrt{\langle\tau\rangle}} \left\{
\left[ \frac{1}{\epsilon} + \log \left( \langle\tau^{mn}\rangle p_{m}p_{n} \right) \right]
I(\tau,p)
-\f16\tau^{ab}\frac{p_{a}p_{b}p_{c}p_{d}}{\langle\tau^{mn}\rangle p_{m}p_{n}}\tau^{cd}
\right\}
$$
\begin{equation}\label{a}
I(\tau,p) = \tau^{ab} \Big[ -\f16 p_{a}p_{b}\langle\tau^{-1}_{cd}\rangle-\f1{24}\langle\tau^{mn}\rangle p_{m}p_{n}
\left( \langle\tau^{-1}_{ab}\rangle\langle\tau^{-1}_{cd}\rangle+2\langle\tau^{-1}_{ac}\rangle\langle\tau^{-1}_{bd}\rangle \right)
+\f16 p_{a}p_{c}\langle\tau^{-1}_{bd}\rangle \Big] \tau^{cd}
\end{equation}
(A modified minimal subtraction scheme has been implemented by adding a finite number to $1/\epsilon$.)


\section{Manifestly covariant effective action}

We can obtain part of the full effective action by promoting the vev's $\langle\tau^{mn}\rangle$ and $\langle\tau^{-1}_{mn}\rangle$ to the full field,
using symmetry principles such as coordinate invariance.  First we write
\begin{equation}\label{aa}
\tau^{mn}=\frac{\sqrt{\gamma}\gamma^{mn}}{(x_5)^2}
\end{equation}
$\gamma^{mn}$ is like a second world-sheet metric (in addition to $g^{mn}$, but Euclidean instead of Minkowskian), with $\gamma=det\ \gamma_{mn}$.
(In D=2 this is a separation of $\tau$ into its determinant and determinant-free parts.)
This introducs an extra degree of freedom that can be gauged away by a local scale invariance:
\[
\gamma^{mn}\rightarrow
\rho^{2}\gamma^{mn}\quad x_{5}\rightarrow\rho^{\frac{2-D}{2}}x_{5}
\]
Since this symmetry holds for arbitrary dimensions, the effective action should still retain this symmetry. Furthermore the one-loop action should be zero degree in
$x_{5}$, since it can be seen as counting the number of loops.

These two requirements allow the following two terms:
\[
A\frac{1}{\epsilon}\sqrt{\gamma}\ x_{5}^{\frac{2-3D}{D-2}}\left[x_{5}^{\frac{4}{D-2}}
\left( \frac{D-2}{4(D-1)}R_{\gamma}-\bo_\gamma \right) \right]^{D/2}x_{5}
\]
and a pure ``gravity" term (independent of $x_{5}$)
\[
B\frac{1}{\epsilon}\sqrt{\gamma}\left(R_{\gamma}\frac{1}{R_{\gamma}-4\frac{D-1}{D-2}\bo_\gamma}R_{\gamma}-R_{\gamma}\right)
\]
where $\bo_\gamma=\frac{1}{\sqrt{\gamma}}\partial_{m}\sqrt{\gamma}\gamma^{mn}\partial_{n}$. Plugging in $D=2+2\epsilon$ we have
\begin{equation}\label{b}
-\sqrt{\gamma} \left\{
A x_{5}^{-1} \left[ \frac1\epsilon + \log( -\bo_\gamma ) \right] \bo_\gamma x_{5}
+B \left( \frac{1}{\epsilon}R_{\gamma} + \ha R_{\gamma}\frac{1}{\bo_\gamma}R_{\gamma} \right)
\right\}
\end{equation}

The coefficients $A$ and $B$ can be determined by comparing to the previous quad\-ratic expansion (\ref{a}).
We express $\tau$ in terms of $x_5$ and $\gamma$, and expand both about their vev's:
\[
x_{5}=\langle x_{5}\rangle+\tilde{x}_{5}
\]
and similarly for $\gamma$.
The $\tilde x_5$-$\tilde\gamma$ crossterm cancels, as expected from (linearized) coordinate invariance.  The $(\tilde x_5)^2$ term is
\begin{equation}
\frac{\langle\sqrt\gamma\rangle}{\langle x_5\rangle^2} \left\{
2\tilde{x}_{5} \left[ \frac{1}{\epsilon} +
\log \left( - \frac{\bo_{\langle\gamma\rangle}}{\langle x_{5}\rangle^{2}} \right) \right]
\bo_{\langle\gamma\rangle}\tilde{x}_5
+ \f23\tilde{x}_{5} \bo_{\langle\gamma\rangle}\tilde{x}_5
\right\}
\end{equation}
(This is equivalent to coupling $x$ to just a scalar.)
One can then see $A=2$.   (The last term is finite and local, and so is regularization dependent, and can be canceled by a finite renormalization.  The same applies to the $\log\langle x_5\rangle$ term.)
Similarly, from the $(\tilde\gamma)^2$ term one finds $B=-1/3$.
(This is equivalent to the usual calculation in a background consisting of just a metric.)  Thus the final form of this part of the bare effective action is
\begin{equation}\label{c}
\sqrt{\gamma} \left\{
-2 x_{5}^{-1} \left[ \frac1\epsilon + \log( -\bo_\gamma ) \right] \bo_\gamma x_{5}
+\f13 \left( \frac{1}{\epsilon}R_{\gamma} + \ha R_{\gamma}\frac{1}{\bo_\gamma}R_{\gamma} \right)
\right\}
\end{equation}

If one tries to convert the above action into $\tau$, using $\tau^{-1/4}=x_{5}$ and $\sqrt{\gamma}\gamma^{mn}=\tau^{mn}/\sqrt{\tau}$, one immediately arrives at the difficulty of rewriting terms depending only on $\gamma^{mn}$, since it is impossible to express it in terms of $\tau$. Furthermore, renormalization of the action (\ref{c}) spoils the scale invariance the unrenormalized effective action was proclaimed to preserve! This is not a surprise, since the pure gravity term (in terms of metric $\gamma^{mn}$, not the world-sheet metric $g^{mn}$) is the usual 2D gravity effective action, which is known to have a conformal anomaly after renormalization. We discuss these difficulties in the next section, and show that one must include the Liouville mode to restore covariance.

\section{Renormalization}

The appearance of a scale anomaly in the ``$B$" term is clear, since it has the same form as the usual gravitational effective action except for the replacement of $g$ with $\gamma$.    The unrenormalized effective action is locally scale invariant by construction, but the infinite, local counterterm breaks the invariance, leaving the renormalized effective action (their difference) anomalous.
The origin of the anomaly in the ``$A$" term is even simpler:  It works in the same way as the scale anomaly for massless matter fields.  (In this case, the analog is $x_5$.)
At $D=2+2\epsilon$ under scaling that term in (\ref{b}) becomes
\begin{equation}
-\sqrt{\gamma}
A x_{5}^{-1} \left[ \frac1\epsilon \left( 1 - 2\epsilon \log\rho \right)
+ \log( - \rho^2\bo_\gamma ) \right] \bo_\gamma x_{5}
\end{equation}
which is indeed invariant. Note the second term comes from the $\epsilon$ piece in $\sqrt{\gamma}'=\sqrt{\gamma}\rho^{-2-2\epsilon}$, which is not present in D=2.

However, unlike the usual scale anomaly for $g$, which re-introduces the scale of the metric as a physical Liouville mode, the scale anomaly for $\gamma$ is a fiction, since $\gamma$ was introduced only as a change of variables from $\tau$.  This second anomaly can be avoided by using the original Liouville mode of $g$ in its place.

The procedure is to scale $\gamma$ by a quantity that will eliminate its anomaly while preserving all physical properties.  As seen above, since the unrenormalized effective action is scale invariant, the only effect will be to add a finite, local counterterm to the renormalized effective action.

A similar problem appears in the expression (\ref{a}) for the two-point function in an arbitrary constant background:  There, instead of $\gamma$ we find $\tau$, which has instead the problem that it breaks coordinate invariance because $\tau^{mn}$ is a density.  But $\tau$ is a scaling of $\gamma$, so the solution is the same.  (In fact, we already needed finite counterterms to relate (\ref{a}) to (\ref{c}).)

Thus the conditions the argument of the log should satisfy with the scaled version of $\gamma$ (or $\tau$) are:
(1) dependence on $\gamma$ only through $\tau$ (i.e., $\gamma$-scale invariance),
(2) degree zero in $\tau$ (since it counts the number of loops), or equivalently space-time dimensionlessness (since only $\tau$ and $x$ carry this dimension), and
(3) coordinate covariance, or equivalently world-sheet dimensionlessness (global scale invariance is a particular coordinate transformation).

Since $\gamma^{mn}$ is itself a scaling of $\tau^{mn}$, the only available quantities with which to scale $\tau^{mn}$ are the determinants of $\tau^{mn}$ and $g_{mn}$, thus satisfying condition (1).  Since the determinant of $g$ is required, its Liouville mode is necessarily introduced.  Condition (2) is then satisfied by multiplying $\tau^{mn}$ by an appropriate power of its determinant, while (3) is satisfied by multiplying by an appropriate power of $g$'s determinant.  This procedure also results in a rescaling of $x_5$, as easily obtained by noting that $\tau$, as expressed in terms of $\gamma$ and $x_5$, is invariant under a rescaling of $\gamma$ and $x_5$ by definition.  The result in arbitrary dimensions is then
$$
\gamma^{mn} \to \frac{\tau^{mn}}{(\sqrt{\tau}\sqrt{-g})^{2/D}}, \qquad
x_5 \to \sqrt{\tau}^{-\frac1D}\sqrt{-g}^{\frac{D-2}{2D}}
$$
Note that now $\sqrt{\gamma}=\sqrt{-g}$, so we have effectively separated the determinant of $\tau^{mn}$ and its unit-determinant part into $x_5$ and $\gamma^{mn}$.

This substitution can be applied to fix the unrenormalized effective action (\ref{c}), but it's simpler to apply directly to the renormalized one, since its net affect is just the addition of finite counterterms to restore the above properties.  Then the final result for covariantly renormalizing (\ref{c}) is
\begin{equation}\label{d}
\Gamma_R[\tau] = \sqrt{\gamma} \left[
-2 x_{5}^{-1} \log \left( - \frac{\bo_\gamma}{m^2} \right) \bo_\gamma x_{5}
+\f16 R_{\gamma}\frac{1}{\bo_\gamma}R_{\gamma}
\right]
\end{equation}
where $m$ is the renormalization scale and
$$ x_5 = \tau^{-1/4}, \qquad \gamma^{mn} \to \frac{\tau^{mn}}{\sqrt{\tau}\sqrt{-g}} $$
or we can simply treat $x_5$ and $\gamma^{mn}$ as new fields replacing $\tau^{mn}$, with the constraint
$$ \sqrt\gamma = \sqrt{-g} $$

The final renormalized action written in the component fields is
\begin{equation}\label{e}
L = \ha\frac{\sqrt{\gamma}\gamma^{mn}\partial_{m}x_{i}\partial_{n}x^{i}+r^{2}\sqrt{\gamma}\gamma^{mn}\partial_{m}x_{5}\partial_{n}x_{5}}{(x_5)^2}+\lambda\tau^{mn}g_{mn}+L_{g}\end{equation}
(or we can replace $\lambda\tau^{mn}g_{mn}$ with $\lambda\gamma^{mn}g_{mn}$), where $r^{2}$ corresponds to the log term in $\Gamma_R$, and the renormalization-invariant mass scale resulting from dimensional transmutation is
$$ M^2 = m^2 e^{-r^2/4} $$
so that the coupling $1/r^2$ is asymptotically free.
(There could also be an $R_\gamma$ term, but it's topological and hence the same as an addition to the $R$ term in $L_g$.)  At this point the only breaking of global scale invariance in the effective action is through the log term, with its scale $M^2$, and the cosmological term, with its scale $\Lambda$.  Thus, the Liouville mode can always be redefined by a constant scale so that these constants appear only through the combination $\Lambda/M^2$, which gives the coupling of the parton theory.  (In fact, without this quantum effect, $\Lambda$ could be scaled away.)

One can choose the coordinate gauge $\sqrt\gamma\gamma^{mn}=\delta^{mn}$; then the Lagrange multiplier enforces the constraint $\delta^{mn}g_{mn}=0$, leaving only two components in the world-sheet metric:  One will be the Liouville mode, contributing a factor of $\sqrt{g}$ that sets the scale for the running.

\section{\texorpdfstring{AdS${}_5$ geometry}{AdS5 geometry}}

The first term in (\ref{e}) looks like the metric for $AdS_{5}$. Indeed the
$AdS_{5}$ metric
\begin{equation}
ds^{2}=r^{2}\frac{(dx_{a})^{2}+(dx_{5})^{2}}{(x_5)^2}
\end{equation}
can be transformed into that of (\ref{e}) by the rescaling
$x_{5}=x_{5}'r^{2}$ so that the metric is
\begin{equation}\label{f}
ds^{2}=\frac{(dx_{a})^{2}+r^{2}(dx_{5}')^{2}}{(x_5')^2}
\end{equation}
In \cite{Nastase:2000za} one considers the classical Type IIB string
propagating in $AdS_{5}\otimes S^{5}$ background in the zero-radius limit,
that is, with the metric in (\ref{f}) and taking the $r\rightarrow0$ limit,
which becomes the projective light-cone. It was shown in that limit the
$S^{5}$ shrinks to zero and the fifth dimension of the $AdS_{5}$ becomes
non-dynamic. Taking the random lattice approach the fifth dimension becomes
a Schwinger parameter and the world-sheet has a natural interpretation as a
planar Feynman diagram.  (In the bosonic case, it is a diagram of massless wrong-sign $\phi^{4}$ theory.) The
coupling constant of the field theory was identified as $r\sim Ng^{2}$;
therefore, this limit corresponds to weak coupling of the field
theory.

Here we consider quantum corrections to the (bosonic) continuum world-sheet theory, corresponding to performing loop-momentum (but not Schwinger parameter) integration in the Feynman diagrams of the field theory. (This is the usual first step in evaluating diagrams.)  We see that the
Schwinger parameter generates the fifth dimension, and the $AdS_{5}$ metric
arises. The radius $r$, or the coupling constant for the field theory, runs in
energy above some scale $M$ set by the Liouville field. Recalling the underlying parton theory
is wrong-sign $\phi^{4}$, which is asymptotically free, in the high-energy limit the theory is at weak coupling. This is reflected in the fact that $r\rightarrow 0$ at high energies, and we are forced onto the projective light-cone of the original geometry. On the other hand, as the energy approaches the scale $M$, $r$ becomes large and the perturbative picture is no longer valid. This is in contrast to earlier effective string theories derived from four-dimensional field theories such as Abelian Higgs models \cite{higgs}. In these
theories one expands around a classical long string configuration; then the
conformal anomaly can be expanded in inverse powers of string length. In the
infinite length limit the theory is perfectly conformally invariant at the
quantum level. In our QCD-like string the emergence of a string in $AdS_{5}$ is really a weak-coupling duality in spirit closer to discussions of string bits, where the correspondence of perturbative N=4 SYM (the limit of vanishing 't Hooft coupling) and tensionless IIB string is examined.

Note that, since the AdS radius $r$ is really a $\log\bo$, some of the isometry of the usual AdS$_{5}$ metric is broken. One can see that the transformations that mix $x_5$ with $x_a$ (these are the conformal boosts) no longer preserve the action. This is not a surprise since the underlying $\phi^{4}$ is not strictly conformal due to the running of the coupling.

An interesting extension of this is the twistor string. In \cite{witten} it
was shown that twistor strings are dual to perturbative N=4 SYM in
4 dimensions at least at tree level. It would seem to imply that the twistor
string is somehow related to the usual type IIB string in the large N limit where the closed-string coupling is suppressed. If one tries to
extract perturbative N=4 SYM from the AdS/CFT
correspondence, it should correspond to a classical string (no closed string
coupling) in the $r^2/\alpha'\rightarrow 0$ limit. This limit can
be taken in two ways: $r^{2}\rightarrow 0$ or $\alpha'\rightarrow \infty$.
The first limit is the limit taken in \cite{Nastase:2000za}, and explicit
calculation of the partition function on both sides seems to agree in this
limit \cite{worldsheetN=4}. The second is taking a tensionless limit. In
\cite{ADHM}, it was shown that the  bosonic part of the ADHM twistor string
is really the tensionless limit of the QCD-like string in \cite{W. Siegel1}.
Thus, all this seems to say that classically the twistor string is the
tensionless limit of the type IIB string. Of course this discussion is really in
the framework of the bosonic part; combining with the fermionic part one encounters the difficulty of rewriting second-class constraints in terms of first-class (perhaps by introducing new gauge symmetry) and consistently reducing the number of $\kappa$ symmetries \cite{tension}.

\section*{Acknowledgements}

This work is supported in part by National Science Foundation
Grant No.\ PHY-0354776.

\end{document}